\documentclass[12pt,a4paper]{article}
\usepackage[utf8]{inputenc}
\usepackage{amsmath}
\usepackage{amsfonts}
\usepackage{amssymb}
\usepackage{makeidx}
\usepackage{graphicx}
\usepackage[left=3cm,right=2cm,top=3cm,bottom=3cm]{geometry}

\title{The Criteria for Quantum Teleportation of an Arbitrary Two-Qubit State Information Based on The Channel Matrices}
\author{Bayu D. Hatmoko, Agus Purwanto, Bintoro Subagyo and Rafika Rahmawati  \\
	~~~\\
	\small Department of Physics, Institut Teknologi Sepuluh Nopember\\ 
	\small Kampus ITS Sukolilo, Surabaya 60111,\\
	\small Indonesia
}

\date{\today}
\begin{document}
	
	\maketitle

\begin{abstract}
We propose channel matrices by using unfolding matrices from their reduced density matrices.  These channel matrices can be a criterion for a channel whether the channel can teleport or not any qubit state. We consider a special case, teleportation of the arbitrary two-qubit state by using the four-qubit channel. The four-qubit channel can only teleport if the rank of the related channel matrix is four.
\end{abstract}

\section{Introduction}
Quantum teleportation is a transmission of the information by using previous shared entangled state and the classical channel between sender and receiver. Quantum teleportation protocol was proposed theoretically by Bennett et al., in 1993 \cite{b.1}. One qubit quantum state is transmitted from Alice to Bob via Einstein-Podolsky-Rosen (EPR)-states as quantum channels. Subsequently, Bouwmeester successfully to demonstrated the experimental of teleportation scheme by using photons as quantum information and channels \cite{b.2}. Then, both theoretical \cite{b.3,b.4,b.5,b.6,b.7, b.8} and experimental \cite{b.9,b.10} research takes much attention in this field.
\par Initially, the one-qubit state was teleported via the channel of two-qubits entangled state. Further development was carried out by Karlsson \cite{b.11} transmitting one-qubit state via the three-qubit channel, i.e GHZ-state. Moreover, Karlsson proposed the third person. In the process, Alice sent one-qubit state to Cliff via Bob as an intermediary. The advantage of the scheme is Cliff as the receiver only apply unitary operations based on Bob's measurement results.  Different from Karlsson, Joo \cite{b.3} investigated two quantum teleportation schemes by utilizing the W-state as a quantum channel. As a result, Joo shows the success of teleporting an unknown qubit via the W-state depends on the type of measurement performed by Alice. However, the teleportation scheme proposed in the reference \cite{b.3,b.11} only consider for sending the one-qubit state.
\par Next, two-qubit quantum teleportation via four-qubit channels was proposed by Rigolin et al. \cite{b.12}. The arbitrary two-qubit states were successfully teleported by using sixteen orthogonal general states constructed from Bell's states as quantum channels. On the other hands, Zha and Song \cite{b.13} expanded the four-qubit quantum channel not only considering the Bell-pair states but also non-bell-pair states as quantum channels. The quantum teleportation can occur if only if Bob's transformation matrix can be determined through the “transformation operator”. Furthermore, Zha and Ren \cite{b.14} analyzed the relationship between "transformation operator" and an invariant under Stochastic Local Operation and Classical Communication (SLOCC). The other approach shows the allowed criteria for quantum teleportation depend on the channel matrices, the measurement matrices, and the collapsed matrices \cite{b.15}.

\par The entanglement is the heart of the quantum teleportation and quantum information in general. For two-qubit state, we just have a simple case that is entangled or separable state. However, higher-qubit cases are more complicated. Dur et al. \cite{b.16} proposed the entanglement class constructed by local unitary operations and classical communication.  In general, the determination of entanglement and separability of multipartite state cannot be represented as a direct product  $\left| \psi \right\rangle = \left| \psi_1 \right\rangle \otimes |\left| \psi_2 \right\rangle \otimes \left| \psi_3 \right\rangle \otimes \cdots$. The separability of the state can be obtained by exploring the rank of the corresponding reduced density matrix. Purwanto et al. \cite{b.17} revealed if the rank of each single reduced density matrix of a multipartite state is one, then the state is completely separable. Contrary, if the rank of each single reduced density matrix is not equal to one, then the type of entangled state shall be determined by investigating various higher dimensional reduced density matrices.

\par In order to teleport an arbitrary two-qubit state via the four-qubit channel, the rank and the entanglement of the channel matrix have to be considered \cite{b.13, b.14, b.15}. In this paper, the reduced density matrix \cite{b.17} are analyzed further to obtain the rank of the four-qubit entangled channel. The density matrix can be expressed as the multiplication of various channel matrices and their Hermitian conjugate. In fact, the channel matrices are nothing but the unfolding matrix \cite{b.18}. 

\par The organization of the paper is as follows. Section 2 describes how the double reduced density matrices can be re-expressed as the multiplication of the channel matrix and its Hermite conjugate. Section 3 demonstrates the implementation of the channel matrix analysis on the teleportation of an arbitrary two-qubit quantum state. The last section serves the conclusions of our work. The unfolding channel matrices for four single reduced density matrices present in the appendix.

\section{Channel matrices}

\par The entanglement of multipartite has been studied by employing the rank of the density matrix \cite{b.17}. If the state of n-partite is completely decomposed, then the rank of every single reduced density matrix is one. Otherwise, if it is not equal to one, the state of n-partite is entangled. However, the entanglement is not necessarily perfect but possibly to be the combination of few entangled sub-states.  In this article, we focus on the state of four-qubit as follows:

\begin{equation}
\left| \phi\right>=\sum_{ijkl=0}^1 c_{ijkl} \left| ijkl\right>, \label{eq.1}
\end{equation}
with the complex coefficient, $C_{ijkl}$ is satisfied by  $|c_{0000} |^2+|c_{0001} |^2+\cdots+|c_{1111} |^2=1$. The density matrix is given by
\begin{equation}
\rho =\left| \phi\right> \left< \phi\right|=\sum_{ijkl}^1\sum_{pqrs}^1 c_{ijkl} c_{pqrs}^{*} \left| ijkl\right> \left< pqrs\right| \label{eq.2}
\end{equation}

which is a $16\times 16$ matrix with rank one. If the state of Eq. (\ref{eq.1}) is completely not entangled, the rank of four single reduced density matrices is one. The rank of all higher reduced density matrices is also equal to one. Since this state cannot teleport, we do not discuss this state further.

\par We are interested in the rank analysis of the entangled state of Eq. (\ref{eq.1}). In such a situation, the rank of four single reduced density matrices is two. Although not specifically discussed, we show the single reduced density matrix in the Appendix. Based on the work of Purwanto et al. \cite{b.17}, we investigate the rank of three double reduced density matrix, i.e., (AB), (AC), and (AD) sub-states. If the rank each of them are not equal to one, the state of Eq. (\ref{eq.1}) is completely entangled. On the other hand, if the rank of one of the doubled reduced density matrices is one, then the state of Eq. (\ref{eq.1}) contain the combination of two entangled states, AB-CD, AC-BD, or AD-BC respectively depend on which rank of the double reduced density matrix is equal to one. The AB doubled sub-states is defined as:
\begin{eqnarray}
\rho_{AB}&=\sum_{mn}^1 \left(I \otimes I \otimes \left< mn \right|\right) \rho \left(I \otimes I \otimes \left| mn \right> \right) \nonumber \\
&=\sum_{ijpq}^1\sum_{mn}^1 c_{ijmn} c_{pqmn}^* \left| ij\right> \left< pq \right|, \label{eq.3}
\end{eqnarray}
which is the $4\times4$ matrix with the index of row $(ij)$ and column $(pq)$. If the rank of this matrix is one, the density matrix rank of its pair (CD) is also one and the state of Eq. (\ref{eq.1}) is combined entangled AB-CD sub-states. However, if the rank is not equal to one, we cannot make any conclusion about the entanglement of state Eq. (\ref{eq.1}) yet. Further investigation of reduced density matrix Eq. (\ref{eq.3}) results
\begin{equation}
\rho_{AB}=C_{AB} C_{AB}^\dagger, \label{eq.4} 
\end{equation}
where $C_{AB}$ is defined as a channel matrix (AB),
\begin{equation}
C_{AB} = \left(\begin{array}{cccc}
c_{0000}&c_{0001}&c_{0010}&c_{0011}\\
c_{0100}&c_{0101}&c_{0110}&c_{0111}\\
c_{1000}&c_{1001}&c_{1010}&c_{1011}\\
c_{1100}&c_{1101}&c_{1110}&c_{1111}
\end{array}\right) \label{5}
\end{equation}
From Eq. (\ref{eq.4}), it is clear that the rank of the density matrix is equal to the rank of the channel matrix. Using this channel matrix, the rank of the density matrix becomes easier to be determined by just considering the channel matrix element. Two others double reduced density matrices are:
\begin{equation}
\rho_{AC}=C_{AC} C_{AC}^\dagger  
\end{equation}
And
\begin{equation}
\rho_{AD}=C_{AD} C_{AD}^\dagger 
\end{equation}
with AC and AD channel matrices as follows:
\begin{equation}
C_{AC} = \left(\begin{array}{cccc}
c_{0000}&c_{0001}&c_{0100}&c_{0101}\\
c_{0010}&c_{0011}&c_{0110}&c_{0111}\\
c_{1000}&c_{1001}&c_{1100}&c_{1101}\\
c_{1010}&c_{1011}&c_{1110}&c_{1111}
\end{array}\right). \label{8}
\end{equation}
\begin{equation}
C_{AD} = \left(\begin{array}{cccc}
c_{0000}&c_{0010}&c_{0100}&c_{0110}\\
c_{0001}&c_{0011}&c_{0101}&c_{0111}\\
c_{1000}&c_{1010}&c_{1100}&c_{1110}\\
c_{1001}&c_{1011}&c_{1101}&c_{1111}
\end{array}\right). \label{9}
\end{equation}
Each of these three channel matrices has sixteen elements in the form of the component of $c_{ijkl}$ from the four-qubit state of Eq. (\ref{eq.1}). The channel matrices in Eq. (\ref{5}), (\ref{8}) and (\ref{9}) is arranged as usual  matrix $C_{IJ}$  in such a way with row $I$-th and column $J$-th are (AB) and (CD), (AC) and (BD) as well as (AD) and (BC) respectively for channel matrices $C_{AB}$, $C_{AC}$ and $C_{AD}$.\\

We consider some of the following examples. First, the GHZ-like state
\begin{equation}
\left|\phi\right> =\frac{1}{\sqrt{2}} \left(\left| 0000\right> + \left| 1111\right>\right), \label{10}
\end{equation}
then each of three-channel matrices has rank two and form
\begin{equation}
C_{AB}=C_{AC}=C_{AD} = \frac{1}{\sqrt{2}} \left(\begin{array}{cccc}
1&0&0&0\\
0&0&0&0\\
0&0&0&0\\
0&0&0&1
\end{array}\right) \label{11}
\end{equation}
It means that the state of Eq. (\ref{10}) is completely entangled since none of the double reduced density matrices has rank one. Moreover, it can be verified easily using the unfolding matrix as given in the Appendix that the rank of all single reduced density matrices is equal to two.\\

The second example is W-like state as given by
\begin{equation}
\left| \phi \right> =\frac{1}{2} \left(\left| 0001\right> + \left| 0010 \right> + \left| 0100\right> + \left| 1000\right>\right), \label{12}
\end{equation}
The sub-states corresponding to Eq. (\ref{12}) then take the form
\begin{equation}
C_{AB}=C_{AC}=C_{AD} = \frac{1}{2} \left(\begin{array}{cccc}
0&1&1&0\\
1&0&0&0\\
1&0&0&0\\
0&0&0&0
\end{array}\right) \label{13}
\end{equation}
One can see that the three-channel matrices have the same form and rank two. Then, the state Eq. (\ref{12}) is completely entangled. Following the same argument for the state Eq. (\ref{10}), the rank of all of its single reduced density matrices is two.\\

The third example is the state
\begin{equation}
\left| \phi \right> =\frac{1}{2} \left(\left| 0000 \right> +\left| 0011 \right> + \left| 1100\right> + \left| 1111\right> \right).  \label{14}
\end{equation}
Subsequently, the corresponding sub-states are
\begin{equation}
C_{AB}=\frac{1}{2} \left(\begin{array}{cccc}
1&0&0&1\\
0&0&0&0\\
0&0&0&0\\
1&0&0&1
\end{array}\right) \label{15}
\end{equation}

\begin{equation}
C_{AC}=C_{AD} = \frac{1}{2} \left(\begin{array}{cccc}
1&0&0&0\\
0&1&0&0\\
0&0&1&0\\
0&0&0&1
\end{array}\right) \label{16}
\end{equation}
The rank of (AB) channel matrix is equal to one while the rank of (AC) and (AD) channel matrix rank is equal to four. It means that the entangled state Eq. (\ref{14}) is the combination of two entangled states, those are AB-CD,
\begin{equation}
\left| \phi \right> =\frac{1}{\sqrt{2}} \left(\left| 00\right> +\left| 11 \right> \right)  \frac{1}{\sqrt{2}} \left(\left| 00\right> +\left| 11 \right> \right). \label{17}
\end{equation}
The fourth example is the state
\begin{equation}
\left| \phi \right> = \frac{1}{2} \left( \left| 0000 \right> + \left| 0011 \right> + \left| 1100 \right> - \left| 1111 \right> \right). \label{18} 
\end{equation}
By comparing the forms of Eq. (\ref{14}) and Eq. (18), as well as channel matrix in Eq. (\ref{15}) and Eq. (\ref{16}), it can be seen that the rank of (AB) channel matrix of the state (\ref{18}) is two while the rank of (AC) and (AD) channel matrices are four. It means, different from the state (\ref{14}), the state (\ref{18}) is completely entangled and known as a cluster \cite{b.19, b.20}.\\

The fifth example is the state
\begin{equation}
\left| \phi \right> = \frac{1}{2} \left(\left| 0000 \right> + \left| 0101 \right> + \left| 1010 \right> + \left| 1111 \right> \right). \label{19}
\end{equation}
Subsequently, the corresponding sub-states take the form
\begin{equation}
C_{AB}=\frac{1}{2} \left(\begin{array}{cccc}
1&0&0&0\\
0&1&0&0\\
0&0&1&0\\
0&0&0&1
\end{array}\right) \label{20}
\end{equation}

\begin{equation}
C_{AC}=\frac{1}{2} \left(\begin{array}{cccc}
1&0&0&1\\
0&0&0&0\\
0&0&0&0\\
1&0&0&1
\end{array}\right) \label{21}
\end{equation}

\begin{equation}
C_{AD}=\frac{1}{2} \left(\begin{array}{cccc}
1&0&0&0\\
0&0&1&0\\
0&1&0&0\\
0&0&0&1
\end{array}\right) \label{22}
\end{equation}
From these three channel matrices, one can obtain that the state of four-qubit (\ref{19}) is AC-BD combined entangled state.\\

The last example is a more elaborate state varied from the state of Eq. (\ref{19}) as
\begin{eqnarray}
\left| \phi \right> &=\frac{1}{4} \left(\left| 0000\right> + \left| 0001 \right> +\left| 0010 \right> +\left| 0011 \right> \right. \nonumber\\
&+ \left| 0100\right> - \left| 0101 \right> +\left| 0110 \right> -\left| 0111\right> \nonumber\\
& + \left| 1000 \right> + \left| 1001 \right> - \left| 1010 \right> -\left| 1011\right> \nonumber\\
&\left.+\left| 1100 \right> - \left| 1101 \right> -\left| 1110\right> +\left| 1111\right> \right). \label{23}
\end{eqnarray}
Subsequently, the corresponding sub-states yield
\begin{equation}
C_{AB}=\frac{1}{4} \left(\begin{array}{rrrr}
1&1&1&1\\
1&-1&1&-1\\
1&1&-1&-1\\
1&-1&-1&1
\end{array}\right) \label{24}
\end{equation}
And
\begin{equation}
C_{AC}=\frac{1}{2} \left(\begin{array}{rrrr}
1&1&1&-1\\
1&1&1&-1\\
1&1&1&-1\\
-1&-1&-1&1
\end{array}\right) \label{25}
\end{equation}

\begin{equation}
C_{AD}=\frac{1}{2} \left(\begin{array}{rrrr}
1&1&1&1\\
1&1&-1&-1\\
1&-1&1&-1\\
1&-1&-1&1
\end{array}\right) \label{26}
\end{equation}
Further evaluation of the three-channel matrices gives the same rank four of (AB) and (AD), while the (AC) channel matrix of rank one. This implies that the state of Eq. (\ref{23}) have the $AC-BD$ pattern of entanglement.

\section{Teleportation of Arbitrary Two-Qubit State}
In this Section, we apply the rank of channel matrix analysis in the previous section on the teleportation of an arbitrary two-qubit state. We consider the state belongs to Alice is given by
\begin{equation}
\left| \chi_A \right>_{12}=\left(x_o \left| 00 \right> +x_1 \left| 01 \right> +x_2 \left| 10 \right> +x_3 \left| 11 \right> \right)_{12}. \label{27}
\end{equation}
As a channel, we adopt the state of four-qubit of Eq. (\ref{eq.1}) with first two-qubit (34) belong to Alice and the other two-qubit (56) belong to Bob,
\begin{equation}
\left| \phi \right>_{3456}=\sum_{ijkl}^1 c_{ijkl} \left| ijkl \right>_{3456}. \label{28}
\end{equation}
There are two other possible qubit combinations belong to Alice, (35) and (36). The combination determines the suitable channel matrices. The main requirement for channel to be able to teleport is that it should be entangled and the entanglement should be between Alice’s qubit and Bob’s qubit, not between Alice’s qubit or Bob’s qubit itself.

Furthermore, Alice join her state to the channel, get
\begin{equation}
\left| \psi\right>_{123456}= \left| \chi_A \right>_{12} \left| \phi\right>_{3456}. \label{29}
\end{equation}
Since the last (56) qubits belong to Bob, Alice’s measurement is applied to the first four qubits (1234) with the projection operator,
\begin{equation}
\left| \mu\right>_{1324}=\sum_{ijkl=0}^1 m_{ijkl} \left| ijkl \right>_{1324}, \label{30}
\end{equation}
with $\sum_{ijkl=0}^1 |m_{ijkl} |^2=1 $. The index (1324) means that the first two qubits of projection operator are projected on the first qubit (first information qubit) and the third qubit (Alice first canal qubit). In other hand, the next two qubits are projected on the second qubit (second information qubit) and the fourth qubit (Alice forth channel qubit). For instance, Alice qubit is (35) then the projection operator become $\left| \mu\right>_{1325}$\\

Following, Alice perform the measurement of Eq. (\ref{29}) by Eq. (\ref{30}) yield,
\begin{equation}
{}_{1324}\left< \mu | \psi\right>_{123456}\equiv \left| \chi \right>_{56}= \sigma_B^{-1} \left| \chi_B \right>_{56} \label{31} 
\end{equation}
The ket $\left| \chi\right>_{56}$ and $\left| \chi_B \right>_{56}$  is Alice’s measurement result and the Bob’s received information respectively. The $\sigma_B^{-1}$ is Bob’s transformation. This equation shows that the teleportation occurs if there is an invertible classical transformation matrix of Bob, $\sigma_B$.  For the measurement based on the Bell and non-Bell state [13], the invertible $\sigma_B$ can be obtained when the channel matrix has rank four. On the other words, teleportation can be held if the rank of the channel matrix is four.

Now, we consider the qubit structure of the channel Eq. (\ref{27}). The qubits (34) belong to Alice and the qubits (56) belong to Bob. In such a configuration, the channel matrix (AB) will determine whether the channel can teleport or not. The teleportation of arbitrary two-qubits information via four-qubits channel can only occur if rank $C_{AB}$  is equal to four. The $C_{AC}$ channel matrix is used to replace $C_{AB}$  when the (35) channel qubit belongs to Alice and the other two channel qubit belong to Bob. The $C_{AD}$ is used in the remaining configuration.

For the measurement based on the state of Bell,

\begin{equation}
\left| \mu \right>_{1324}^{ij}=\left(\left| \beta^i \right> \left| \beta^j \right> \right)_{1324} \label{32}
\end{equation}
with 
\begin{equation}
\left| \beta^{1,2} \right>=\frac{1}{\sqrt{2}} \left(\left| 00 \right> \pm \left| 11 \right> \right), \label{33}
\end{equation}

\begin{equation}
\left| \beta^{3,4} \right>=\frac{1}{\sqrt{2}} \left(\left| 01 \right> \pm \left| 10 \right> \right). \label{44}
\end{equation}
We apply these measurements to the channels state mentioned in section 2. First, we consider the channel of Eq. (\ref{10}) and Eq. (\ref{12}). In this case, the rank of all the channel matrices $C_{AB}$, $C_{AC}$ and $C_{AD}$ are not equal to four but two. Then, both of the channels cannot teleport the state of Eq. (\ref{27}) for all measurement. In other word, no invertible $\sigma_B$ satisfy Eq. (\ref{31}).

Second, the channel of Eq. (\ref{14}) has rank one of $C_{AB}$. However, the rank of $C_{AC}$ and $C_{AD}$ are equal to four. Since Alice’s channel qubit is the first and the second qubit then the appropriate channel matrix is $C_{AB}$. The rank of $C_{AB}$ is not equal to four then the channel can not teleport the state of Eq. (\ref{27}) even though the rank of $C_{AC}$ and $C_{AD}$  are equal to four.

Third, the channel of Eq. (\ref{19}) can teleport since the rank of $C_{AB}$ is equal to four. In Principe, we can obtain invertible $\sigma_B$ for all measurement.  For example, the measurement $\left| \mu \right>_{1324}^{13}$, using Eq. (\ref{31}) yield the rank four of the channel matrix $C_{AB}$  with

\begin{equation}
\sigma_B = 4 \left( \begin{array}{cccc}
0&1&0&0\\
1&0&0&0\\
0&0&0&1\\
0&0&1&0\\
\end{array}\right) \label{35}
\end{equation}

Finally, similar to the previous case, the rank of $C_{AB}$ from the state of Eq. (\ref{23}) is also four. For projection operator $\left| \mu \right>_{1324}^{11}$, we obtain the invertible $\sigma_B$,

\begin{equation}
\sigma_B = 2 \left( \begin{array}{rrrr}
1&1&1&1\\
1&-1&1&-1\\
1&1&-1&-1\\
1&-1&-1&1\\
\end{array}\right) \label{36}
\end{equation}
For the measurement based on the non-Bell state \cite{b.13}:
\begin{equation}
\left| \phi\right> =\frac{1}{2} \left( \left| 0000 \right> + \left| 0101 \right> + \left| 1011 \right> + \left| 1110 \right> \right), \label{37} 
\end{equation}
then, we apply the above measurement of Eq. (\ref{37}) to the channel state in Eq. (\ref{19}) to obtain
\begin{equation}
\sigma_B = 4 \left( \begin{array}{cccc}
1&0&0&0\\
0&1&0&0\\
0&0&0&1\\
0&0&1&0\\
\end{array}\right)
\end{equation}
Furthermore, applying Eq. (\ref{37}) to channel state in Eq. (\ref{23}), we get:
\begin{equation}
\sigma_B = 2 \left( \begin{array}{rrrr}
1&1&1&1\\
1&-1&1&-1\\
1&-1&-1&1\\
1&1&-1&-1\\
\end{array}\right)
\end{equation}
For the measurement based on non-Bell state in Eq. (\ref{37}), and channel state in Eq. (\ref{19}) and (\ref{23}), it is clear that the teleportation is success. 

\section{Conclusion and Outlook}
In this work, we investigate the suitable channel to teleport using the rank of the channel matrix using Bell states as a measurement. The single reduced density matrix of the four-qubit channel state can be expressed in the $2\times8$ unfolding matrices multiplication. The elements of the matrices are in the form of channel state coefficient. In particular, employing the unfolding matrix to the channel state, the rank of single reduced density matrix can be determined easily by regarding the component of channel matrices without any calculation. Moreover, the double reduced density matrices can be expressed by multiplication of $4\times 4$ channel unfolding like matrices. The quantum teleportation of an arbitrary two-qubit state is available if and only if the appropriate $4\times4$ channel unfolding like matrix has rank four.  The method proposed in this work is easier than previous work by Zha \cite{b.13, b.14, b.15}, and we suggest it can be applied in more qubit states, for instance, in \cite{b.6} and \cite{b.7}. 

\section*{Acknowledgement}
This work is supported partially by the Ministry of Research, Technology and Higher Education, Indonesia and BDH thanks to Dr. Lila Yuwana for a fruitful discussion.

\section*{Appendix}

\appendix
\setcounter{section}{1}
A single reduced density matrix of Eq. (\ref{eq.2}) is defined by
\begin{eqnarray}
\rho_A =\sum_{mnt}^1\left(I\otimes \left< mnt\right> \right) \rho \left(I \otimes \left| mnt \right>\right) \nonumber\\
=\sum_{ip}^1 \sum_{mnt}^1 c_{imnt} c_{pmnt}^{*} \left| i \right> \left< p \right| \label{41}
\end{eqnarray}
In a more explicit form, $\rho_A$ is given by
\begin{eqnarray}
\rho_A &= \left(\begin{array}{cc}
\sum_{mnt}^1 c_{0mnt}c_{0mnt}^{*}&\sum_{mnt}^1 c_{0mnt}c_{1mnt}^{*}\\
\sum_{mnt}^1 c_{1mnt}c_{0mnt}^{*}&\sum_{mnt}^1 c_{1mnt}c_{1mnt}^{*}\\
\end{array}\right) \nonumber\\
&= C_A C_A^\dagger
\end{eqnarray}
With
\begin{equation}  \label{A3} 
C_A = \left( \begin{array}{cccccccc}
c_{0000}&c_{0001}&c_{0010}&c_{0011}&c_{0100}&c_{0101}&c_{0110}&c_{0111}\\
c_{1000}&c_{1001}&c_{1010}&c_{1011}&c_{1100}&c_{1101}&c_{1110}&c_{1111}
\end{array}\right)\\ 
\end{equation} 
In a similar manner, the other three single reduced density matrices result
\begin{eqnarray}
\rho_B = C_B C_B^\dagger\\
\rho_C = C_C C_C^\dagger\\
\rho_D = C_D C_D^\dagger
\end{eqnarray}
with
\begin{equation}  \label{A7} 
C_B = \left( \begin{array}{cccccccc}
c_{0000}&c_{1000}&c_{0001}&c_{1001}&c_{0010}&c_{1010}&c_{0011}&c_{1011}\\
c_{0100}&c_{1100}&c_{0101}&c_{1101}&c_{0110}&c_{1110}&c_{0111}&c_{1111}
\end{array}\right)\\
\end{equation} 
\begin{equation}  \label{A8} 
C_C = \left( \begin{array}{cccccccc}
c_{0000}&c_{0100}&c_{1000}&c_{1100}&c_{0001}&c_{0101}&c_{1001}&c_{1101}\\
c_{0010}&c_{0110}&c_{1010}&c_{1110}&c_{0011}&c_{0111}&c_{1011}&c_{1111}
\end{array}\right)
\end{equation}
\begin{equation} \label{A9} 
C_D = \left( \begin{array}{cccccccc}
c_{0000}&c_{0010}&c_{0100}&c_{0110}&c_{1000}&c_{1010}&c_{1100}&c_{1110}\\
c_{0001}&c_{0011}&c_{0101}&c_{0111}&c_{1001}&c_{1011}&c_{1101}&c_{1111}
\end{array}\right)
\end{equation}  
These four matrices $C_A$, $C_B$, $C_C$ and $C_D$  are nothing but unfolding matrices \cite{b.18}.

For example, applying to the channel state of Eq. (\ref{10}), we obtain
\begin{eqnarray}
C_A = C_B = C_C = C_D &\nonumber\\
&= \frac{1}{\sqrt{2}} \left( \begin{array}{cccccccc}
1&0&0&0&0&0&0&0\\
0&0&0&0&0&0&0&1
\end{array} \right) \label{50}
\end{eqnarray}
with rank two. For channel state of Eq. (\ref{8}), we have
\begin{eqnarray}
C_A = C_B = C_C = C_D&\nonumber\\
&= \frac{1}{2} \left( \begin{array}{cccccccc}
0&1&1&0&1&0&0&0\\
1&0&0&0&0&0&0&0
\end{array} \right)\label{51}
\end{eqnarray}
with rank two. We can know the rank of the channel matrices of single reduced density matrix only by easily seeing the component in the matrix of Eq. (\ref{50}-\ref{51}). In the state of Eq. (\ref{14}), we have
\begin{eqnarray}
C_A  = C_C = \frac{1}{2} \left( \begin{array}{cccccccc}
1&0&0&1&0&0&0&0\\
0&0&0&0&1&0&0&1
\end{array} \right) \label{52}
\end{eqnarray}
\begin{eqnarray}
C_B  = C_D = \frac{1}{2} \left( \begin{array}{cccccccc}
1&0&0&0&0&0&1&0\\
0&1&0&0&1&0&0&1
\end{array} \right)  \label{53}
\end{eqnarray}
Similarly, by seeing at a glance the Eq. (\ref{52}-\ref{53}), we know the rank of the channel matrices are equal to two.
Below is the example of non-entangled state,
\begin{equation}
\left| \phi \right> =\frac{1}{2} \left(\left| 0001 \right> + \left| 0011 \right> + \left| 0101 \right> + \left| 0111 \right> \right)  \label{54}
\end{equation}
Then
\begin{equation}
C_A = \frac{1}{2} \left( \begin{array}{cccccccc}
0&1&0&1&0&1&0&1\\
0&0&0&0&0&0&0&0
\end{array}\right)
\end{equation}
\begin{equation}
C_B = \frac{1}{2} \left( \begin{array}{cccccccc}
0&0&1&0&0&0&1&0\\
0&0&1&0&0&0&1&0
\end{array}\right) 
\end{equation}
\begin{equation}
C_C = \frac{1}{2} \left( \begin{array}{cccccccc}
0&0&0&0&1&1&0&0\\
0&0&0&0&1&1&0&0
\end{array}\right)
\end{equation}
\begin{equation}
C_D = \frac{1}{2} \left( \begin{array}{cccccccc}
0&0&0&0&0&0&0&0\\
1&1&1&1&0&0&0&0
\end{array}\right)
\end{equation}
with each of channel matrices rank is equal to one. The state of Eq. (\ref{54}) is decomposed into
\begin{equation}
\left| \phi\right> = \left| 0 \right>  \frac{1}{\sqrt{2}} \left(\left| 0 \right> + \left| 1 \right> \right)  \frac{1}{\sqrt{2}} \left(\left| 0 \right> + \left| 1 \right> \right) \left| 1 \right>. 
\end{equation}

\end{document}